\documentclass[amsmath,amssymb,floatfix,aps,prl]{revtex4-2}
\usepackage{graphicx}
\usepackage{xcolor}
\usepackage{dcolumn}
\usepackage{bm}
\usepackage{amsmath}
\usepackage{subcaption}
\usepackage{caption}
\usepackage{float}
\usepackage{comment}
\usepackage{stackengine} 
\usepackage{adjustbox}

\usepackage{setspace}

\usepackage[normalem]{ulem}
\usepackage{color}
\usepackage{soul}
\setstcolor{red}
\newcommand{\Frac}[2]{\frac{\displaystyle #1}{\displaystyle #2}}
\newcommand{\dup}{{\mathrm d}}
\newcommand{\subtext}[1]{\mbox{\scriptsize #1}}
\newcommand{\dpart}{d_{\subtext{p}}}

\begin{document}
\title{The flow deep within granular piles}
\author{Aqib Khan}
 \author{Prabhu R Nott}%
\affiliation{%
 Department of Chemical Engineering, Indian Institute of Science, Bangalore 560012, India\\}%

\begin{abstract}
Grain piles embody the complex mechanics and kinematics of disordered granular materials, including solid-like and fluid-like behaviours, complex kinematics, and preparation history-dependent stress variation.  It is widely believed that the bulk of a growing pile is static and flow is confined to a thin layer at the surface, but very few studies have investigated the subsurface kinematics.  Here we study the flow within conical grain piles by flow imaging experiments and particle dynamics simulations.  We provide direct evidence of continuous plastic flow deep within piles as grains are poured from above, and show that the direction of flow varies smoothly from vertical at the symmetry axis to parallel to the surface at the periphery.  Our findings provide new insight into the kinematics and rheology of granular media, including the nature of creep in seemingly solid-like regions, and have important implications for geophysical phenomena such as landslides and industrial processes.
\end{abstract}

\maketitle

Piles and heaps of granular materials abound in nature, agriculture and myriad industries, and illustrate several distinctive features of the mechanics and kinematics of granular media \cite{nedderman-book,rao_nott,andreotti_etal-book}.  When grains are deposited on top of a conical pile, they appear to simply flow down the surface.  Indeed, it is widely believed that flow in a pile is confined to a thin surface layer \cite{,jaeger_etal1996,khakhar_etal2001,andreotti_douady2001,isner_etal2020}, and that the pile grows at constant shape through accretion of grains to its static core \cite{wittmer_etal1996, savage1998}.  The persistence of a static pile after grain deposition is stopped lends support to the assumption of surface flow. However, certain predictions of continuum models and some experimental observations of the mechanical response cast doubt on the assumption.  Continuum plasticity theories describe the coexistence of static and flowing regions through a stress-based yield criterion, but commonly used yield criteria are satisfied throughout the pile, not just at the surface \cite{nedderman-book,rao_nott}.  Secondly, experiments show that the stress distribution at the base of a pile depends on whether it is formed by pouring grains through a narrow funnel or by raining them from a distributed source \cite{jotaki_moriyama1979, smid_novosad1981,krishnaraj_nott2024}, but it is unclear how such different stress states can arise if the flow is confined to the surface, and therefore similar, in both cases.  On the other hand, observations in geophysical settings suggest that small disturbances in granular masses on hillslopes can lead to deformation deep within the medium, sometimes leading to catastrophic events such as landslides and avalanches~\cite{roering_etal2001, ferdowsi_etal2018, deshpande_etal2021}

A few studies have investigated the flow at the surface of conical piles and have characterized it as a thin layer of grains flowing down the slope~\cite{altshuler_etal2003,isner_etal2020} in the rapid flow (grain inertia) regime.  The surface flow in quasi two dimensional piles, where a layer of granular material is confined between transparent vertical plates, has been better investigated \cite{khakhar_etal2001, komatsu_etal2001, andreotti_douady2001, crassous_etal2008, martinez_etal2016}; some of the studies have found a sublayer of slow, non-inertial flow where the velocity decays exponentially with depth \cite{komatsu_etal2001, crassous_etal2008, martinez_etal2016}, beneath the inertial surface layer.  A related problem that has been studied is the flow on the surface of partially filled, horizontal rotating drums, where the measurements are made near a transparent end cap; here too the flow is characterized as occurring in a thin surface layer above a core that rotates as a solid body~\cite{orpe_khakhar2001,jain_etal2002}.  A limitation of the experiments in quasi 2D piles and drums is that the velocities measured in them are likely to be substantially smaller than in the bulk, due to the retarding effect of the confining walls~\cite{jop_etal2005}.  Moreover, all the studies cited above have measured only the velocity parallel to the surface.  A recent computational study based on particle dynamics simulations \cite{krishnaraj_nott2024} has revealed the presence of a more complex internal flow within a conical pile during its formation, but the velocity field was not fully characterized. To our knowledge there are no reported experimental investigations of flow deep within the bulk, primarily due to the opacity of granular materials and the limitations of non-invasive imaging techniques.  Investigating the subsurface flow in grain piles will improve understanding of the kinematics of granular media in configurations where solid-like and fluid-like regions coexist, and assist in building a robust continuum description.  It will also be useful in deciphering geophysical processes and advancing industrial applications.

In this paper we investigate the flow in free-standing conical grain piles.  We conduct experiments on steady flowing piles by pouring grains at a constant rate onto the top of a pile resting on a transparent base and imaging the flow at the base from below. We complement our experiments with particle dynamics simulations, which provide the velocity, stress and density fields throughout the pile.  Our study presents compelling evidence for the presence of flow throughout the pile, challenging the conventional notion that most of the pile is static.  Our data for various pile sizes and deposition rates yields a rough scaling of the velocity at the base with the deposition velocity and pile size, which suggests that the velocity magnitude decreases as the inverse-square of the pile diameter.

\begin{figure}
    \captionsetup{justification=raggedright,singlelinecheck=false}
    \begin{minipage}{0.3\textwidth}
     \begin{subfigure}{\textwidth}
      \caption{} \vspace*{-0.5em}
        \includegraphics[width=\textwidth,trim={0.7cm 0.2cm 1.4cm 0cm},clip]{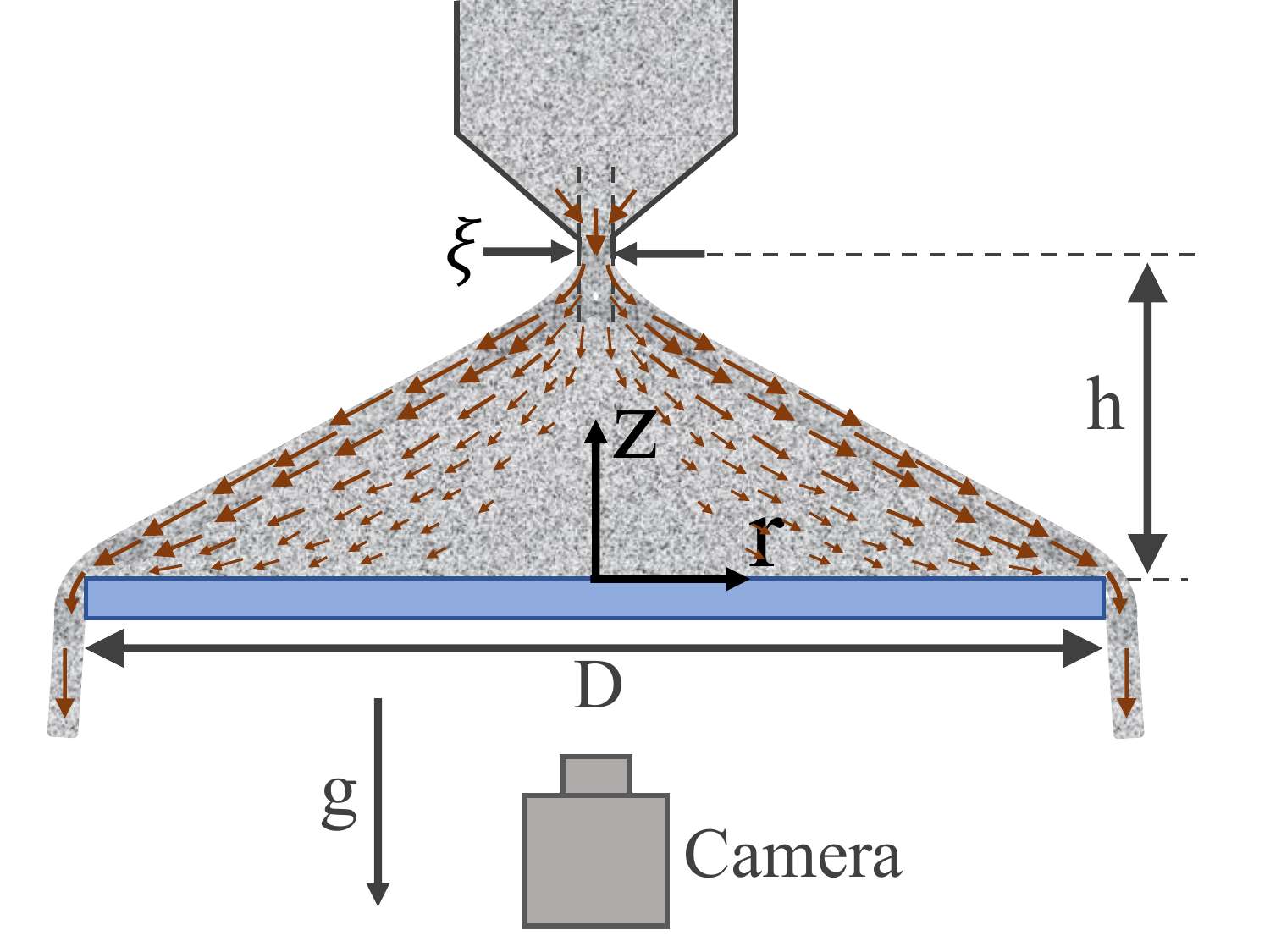} 
       \end{subfigure}  
    \end{minipage}%
    \hspace{0.05\textwidth}  
    \begin{minipage}{0.28\textwidth}
        \begin{subfigure}{\textwidth}
        \caption{ } \vspace*{-0.5em}
            \includegraphics[width=\linewidth,trim={0.5cm 5.2cm 0.26cm 4.0cm},clip]{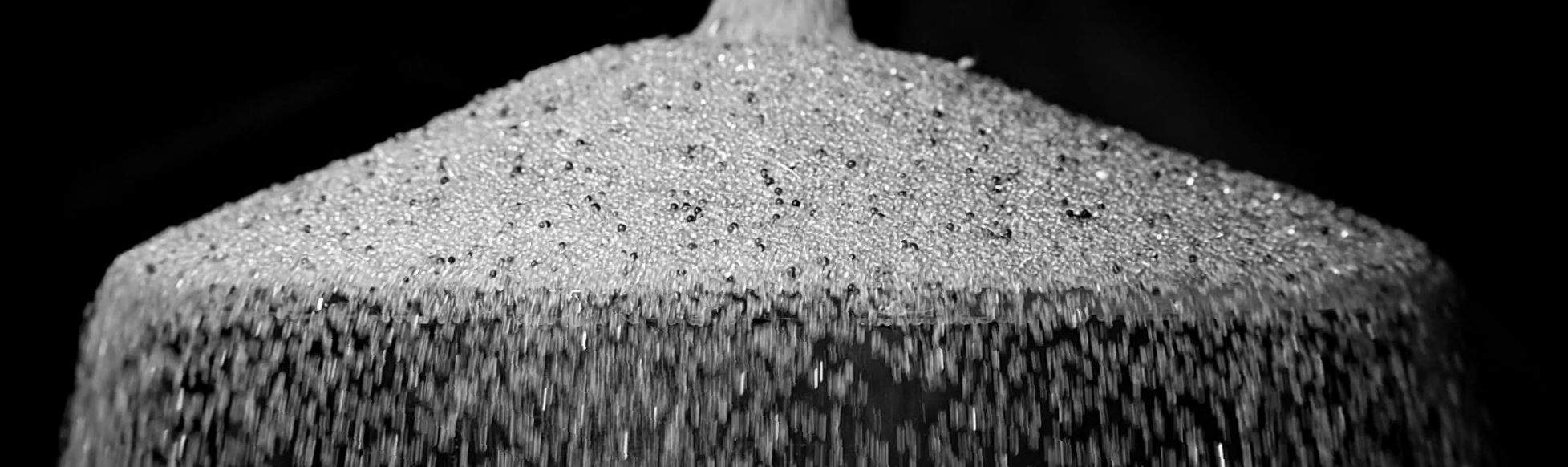} \vspace*{-0.04cm}
        \end{subfigure}
       \vspace{0.04cm} 
        \begin{subfigure}{\textwidth}
        \caption{ } \vspace*{-0.5em}
            \includegraphics[width=\linewidth,trim={0.0cm 0cm 0.0cm 0cm},clip]{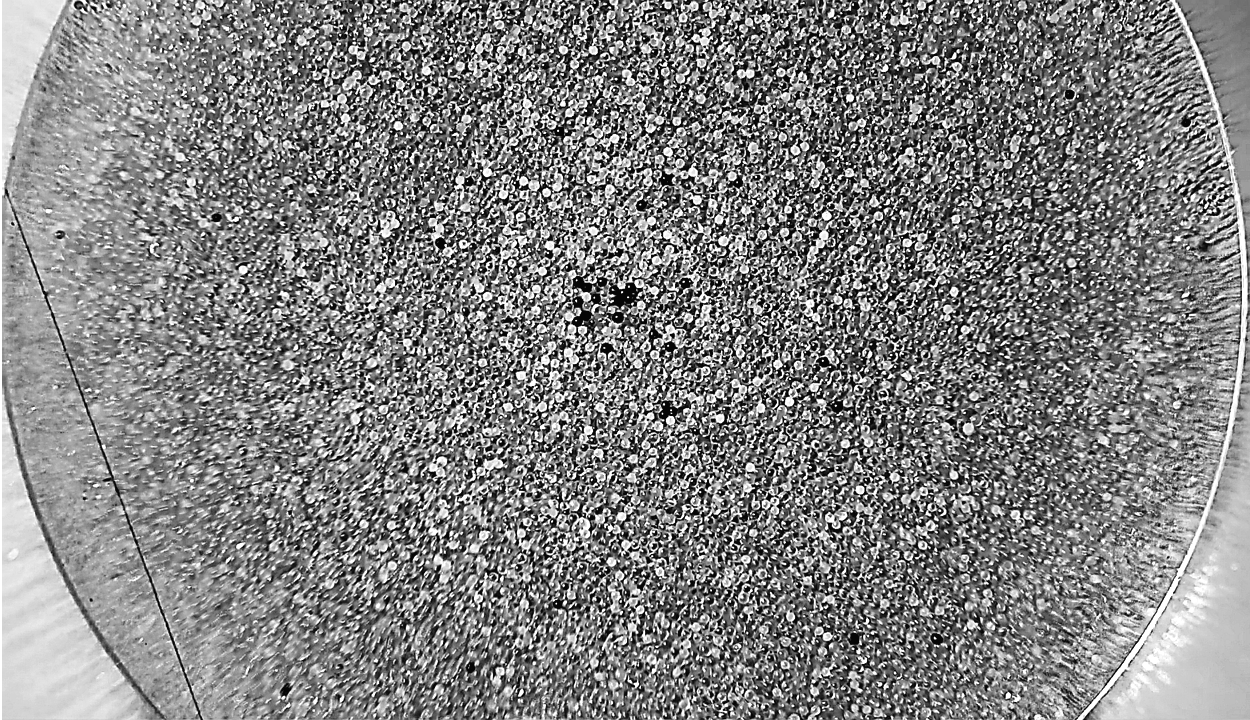} 
             
        \end{subfigure}
    \end{minipage}
    
\caption{(a) Schematic of the experimental setup.  The video camera images the base of the pile through the transparent plate.  (b) Snapshot of a steady-flowing pile of glass beads from the side.  At steady state, the rate of efflux of grains from the hopper equals the rate at which grains spill over from the base.  (c) Snapshot of a steady-flowing pile taken by the video camera below the base.  The images recorded by the camera were analyzed to obtain the radial velocity profile at the base (see Methods).\label{fig1}}
\end{figure}

In our experiments, grains of mean diameter $\dpart$ were dispensed from a hopper onto a horizontal, transparent acrylic disk of diameter $D$ (Fig.~\ref{fig1}a) to form a conical pile.  The flow from the hopper was continued and grains allowed to flow out from the periphery of the disk (Figs~\ref{fig1}a,b) to achieve a steady flowing state; all the measurements were made in this steady state.  We focused on the steady state in order to increase the duration over which flow imaging was conducted, thereby getting better statistical sampling and improving accuracy of the velocity measurements.  The hopper orifice was positioned just slightly above the apex of the pile so as to minimize the impact of the grains on the pile.  A video camera positioned below the transparent disk imaged the base of the pile (Fig.~\ref{fig1}c); the velocity field at the base was determined from the movies by particle image velocimetry.  Piles of different sizes were created by varying the diameter $D$ of the disk, and the mass flow rate $Q$ from the hopper was varied by varying the diameter $\xi$ of the orifice.  We studied piles formed with glass beads, mustard seeds and beach sand.  Further details of the experimental assembly, protocol, materials, and measurement techniques given in Methods.

\begin{figure*}[t]
\centering
$\begin{array}{c c c c c}
\hspace{-0.12cm}
\stackinset{l}{35pt}{t}{-10pt}{\textcolor{black}{$t = 0\,$s}}{
            \includegraphics[width=0.197\textwidth,trim={0cm 0cm 0cm 0cm},clip]{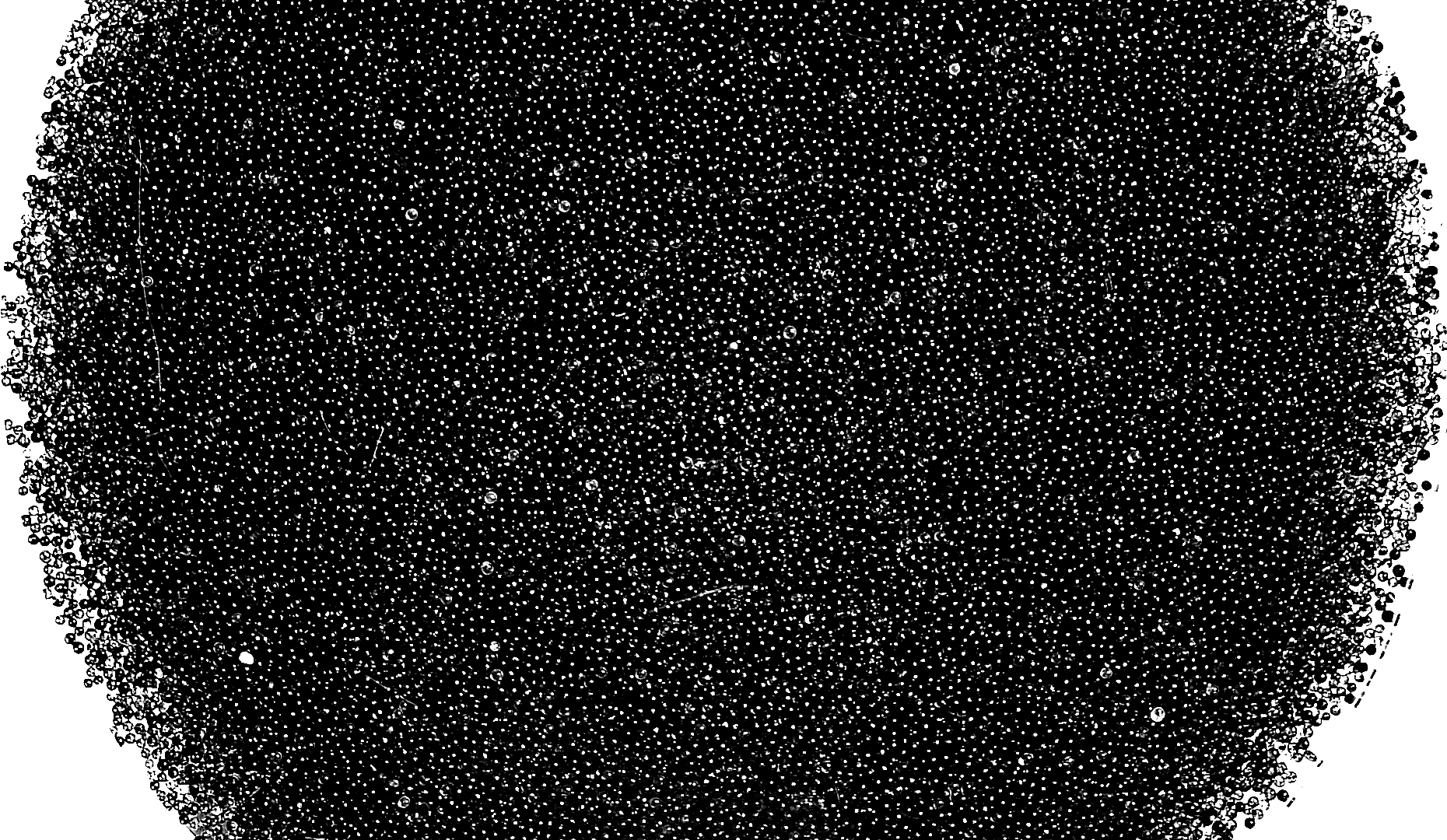}
        }

&
\hspace{-0.40cm} 
\stackinset{l}{35pt}{t}{-10pt}{\textcolor{black}{$t = 30\,$s}}{
\includegraphics[width=0.197\textwidth,trim={0cm 0cm 0cm 0cm},clip]{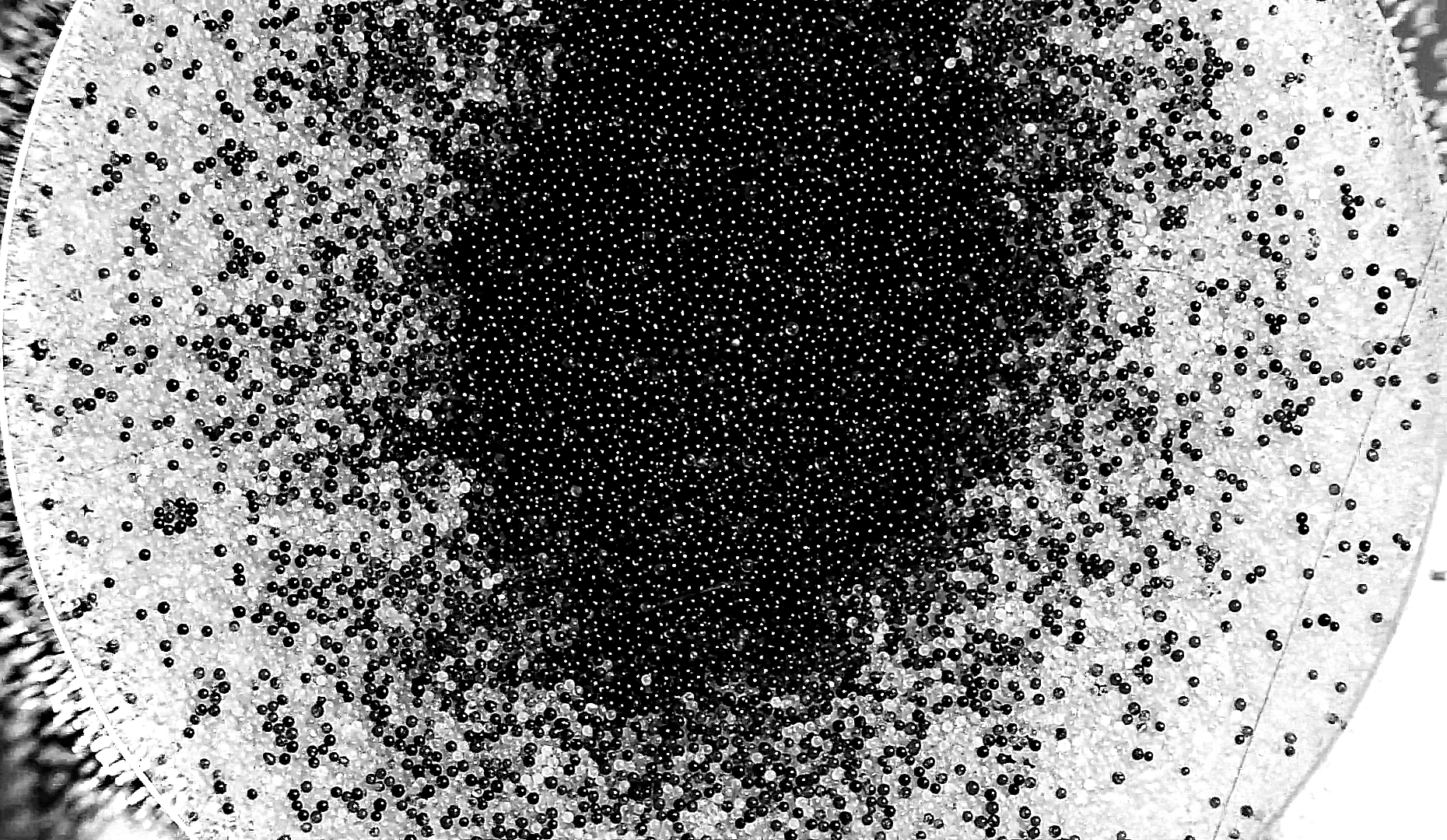} 
}
&
\hspace{-0.285cm} 
\stackinset{l}{35pt}{t}{-10pt}{\textcolor{black}{$t = 120\,$s}}{\includegraphics[width=0.197\textwidth,trim={0cm 0cm 0cm 0cm},clip]{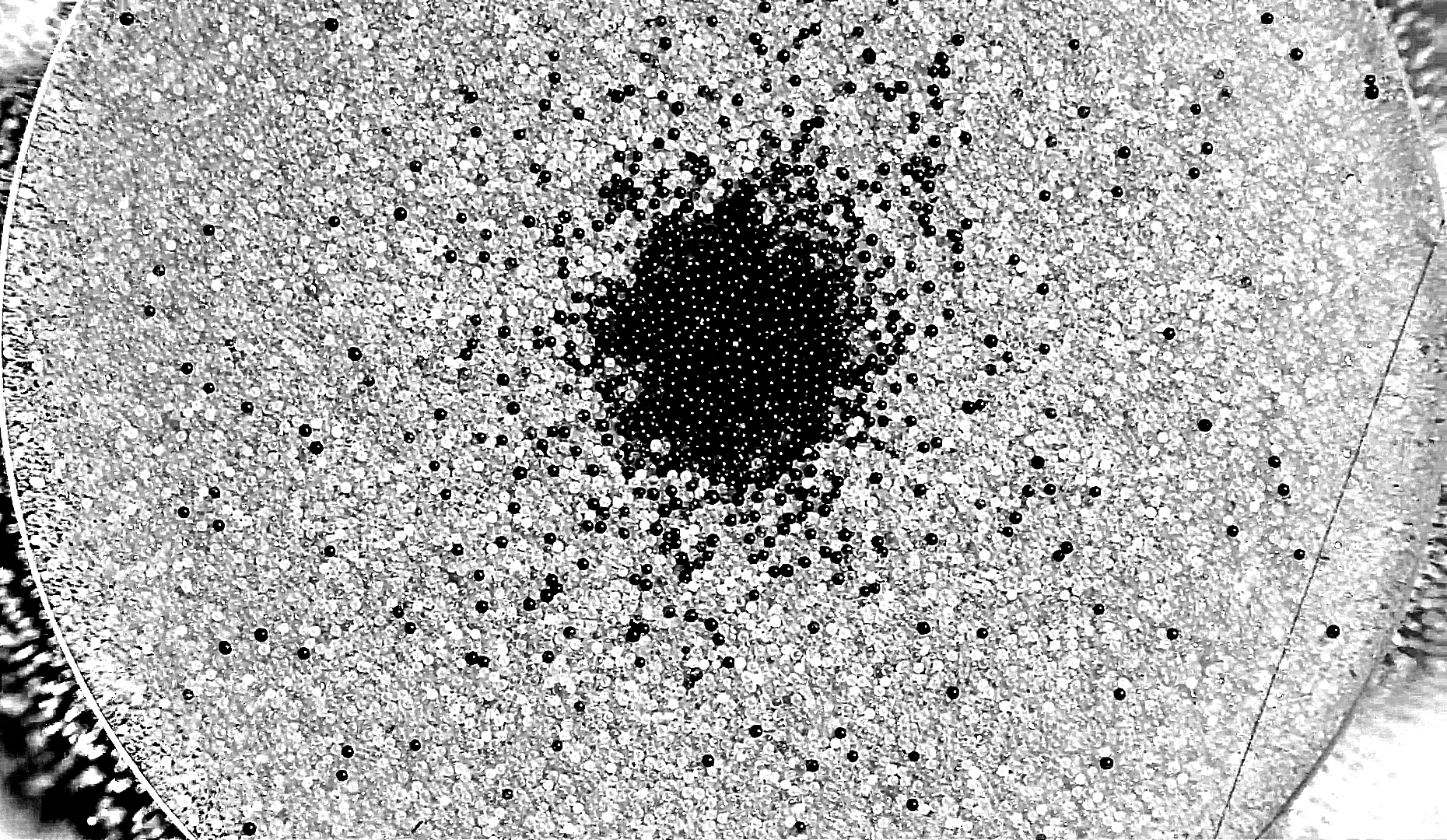} 
}
&
\hspace{-0.40cm} 
\stackinset{l}{35pt}{t}{-10pt}{\textcolor{black}{$t = 300\,$s}}{
\includegraphics[width=0.197\textwidth,trim={0cm 0cm 0cm 0cm},clip]{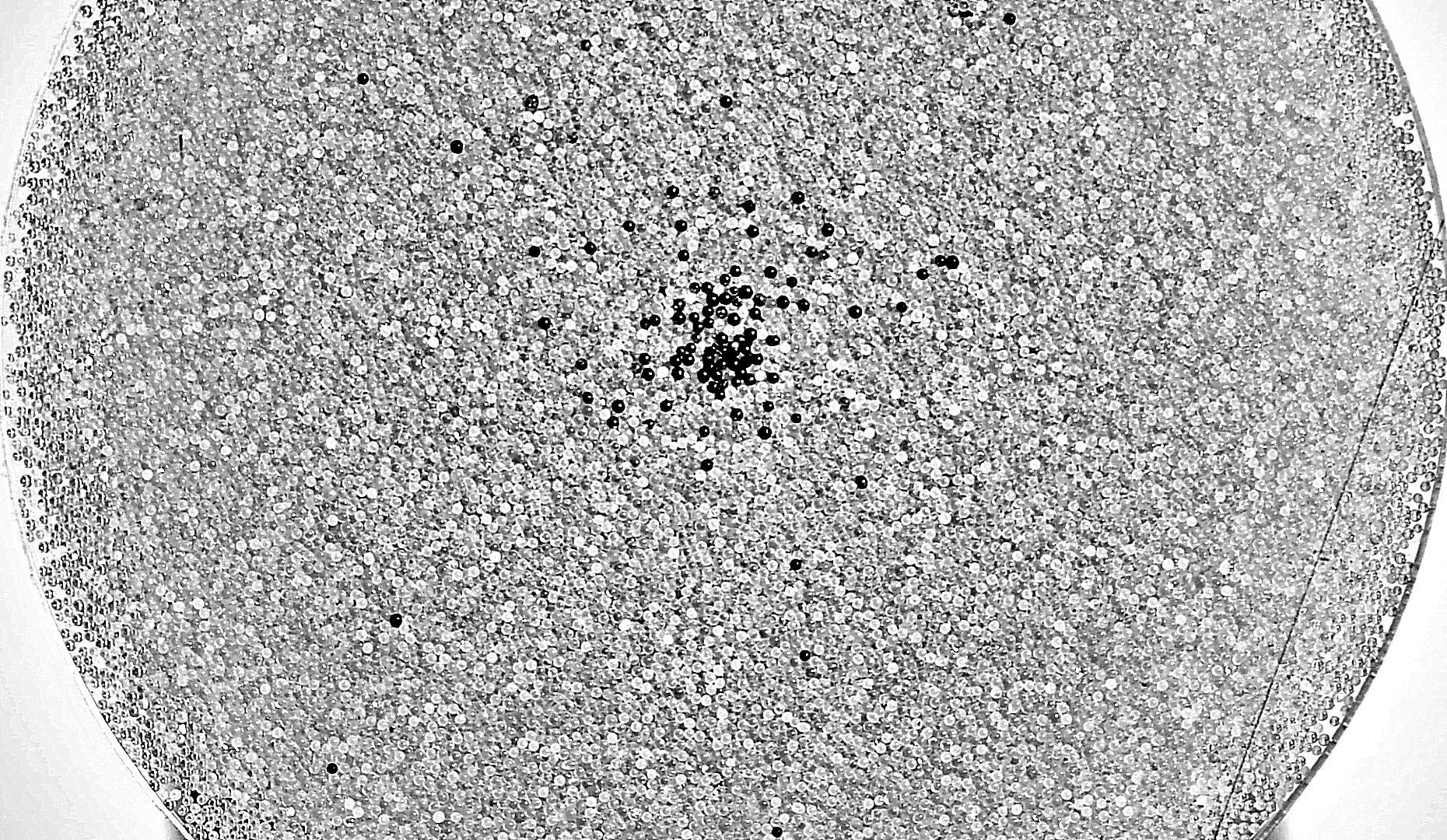}
}
&
\hspace{-0.40cm} 
\stackinset{l}{35pt}{t}{-10pt}{\textcolor{black}{$t = 600\,$s}}{
\includegraphics[width=0.197\textwidth,trim={0cm 0cm 0cm 0cm},clip]{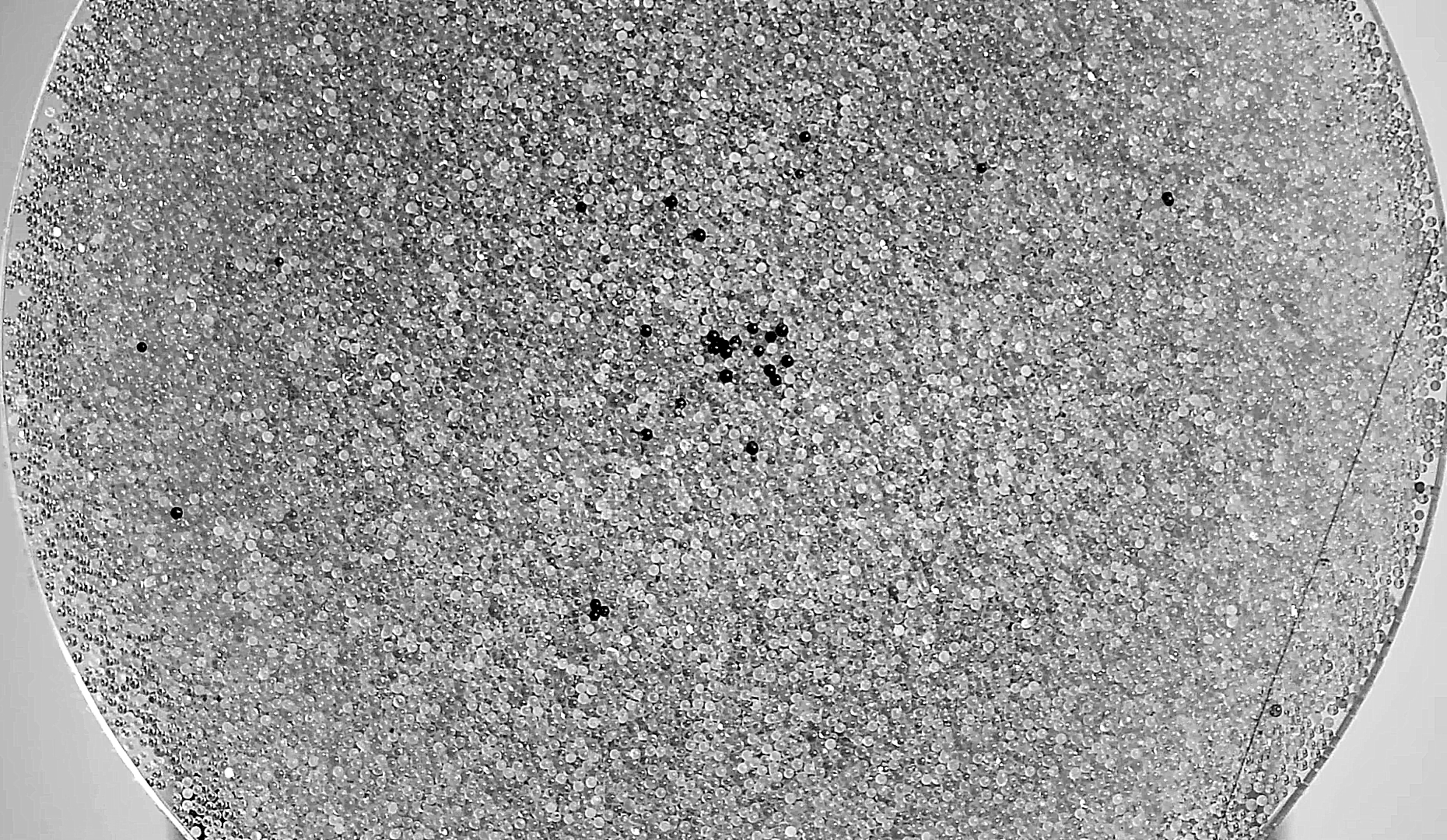}
}
\end{array}$

$\begin{array}{c c c c c}
\includegraphics[width=0.197\textwidth,trim={0cm 0cm 0cm 0cm},clip]{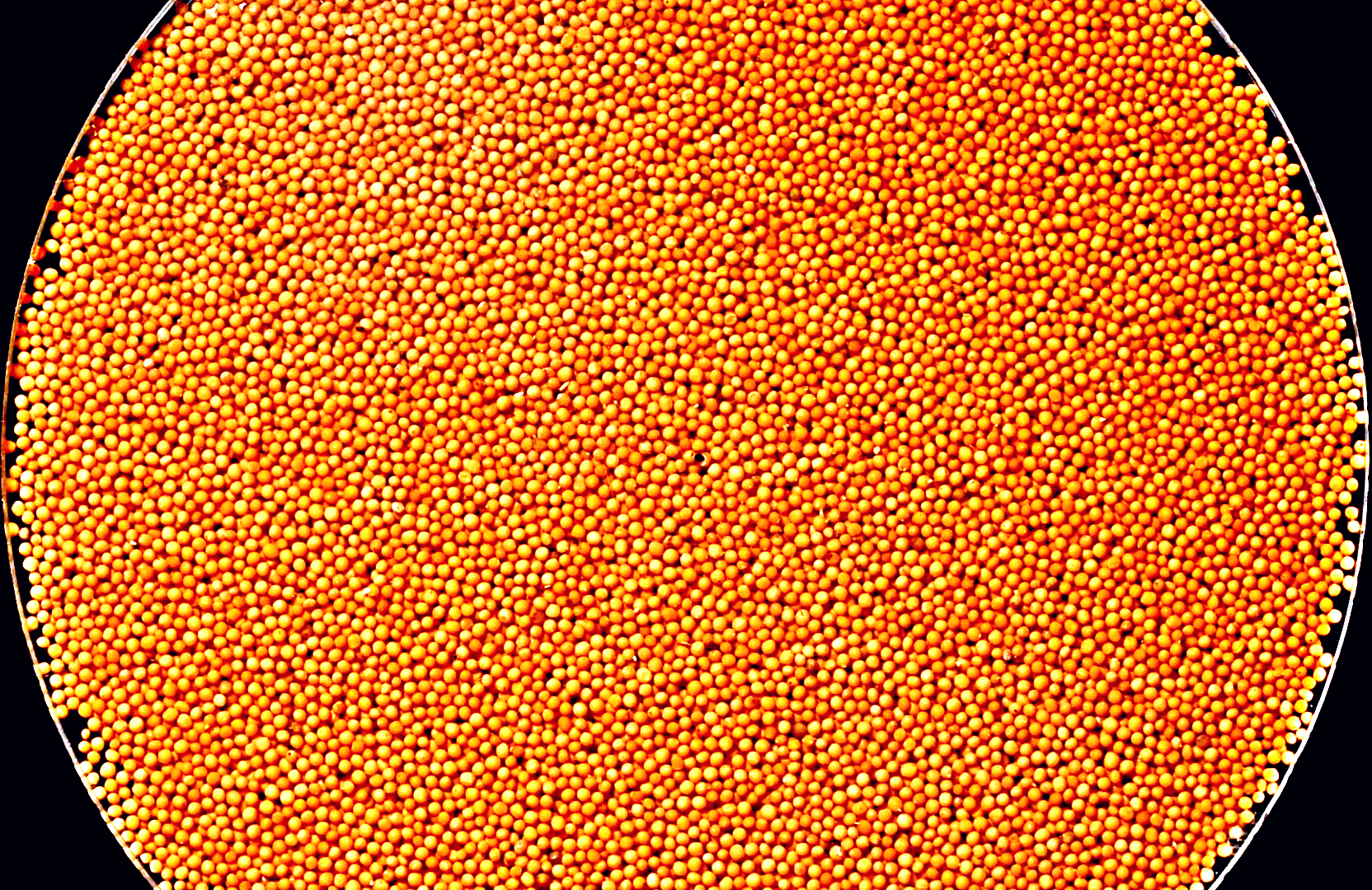}
&
\hspace{-0.17cm} \includegraphics[width=0.197\textwidth,trim={0cm 0cm 0cm 0cm},clip]{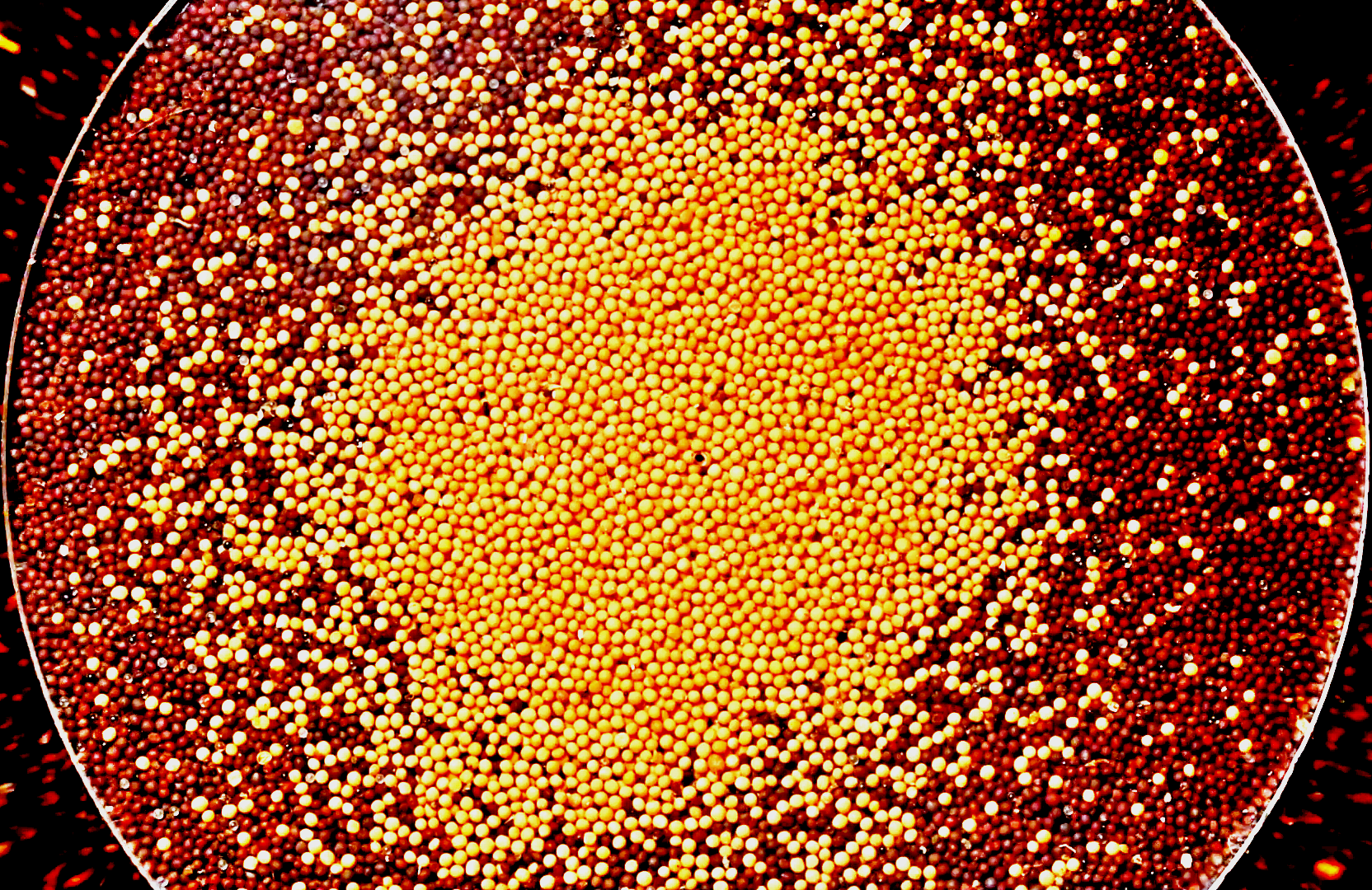}
&
\hspace{-0.17cm} \includegraphics[width=0.197\textwidth,trim={0cm 0cm 0cm 0cm},clip]{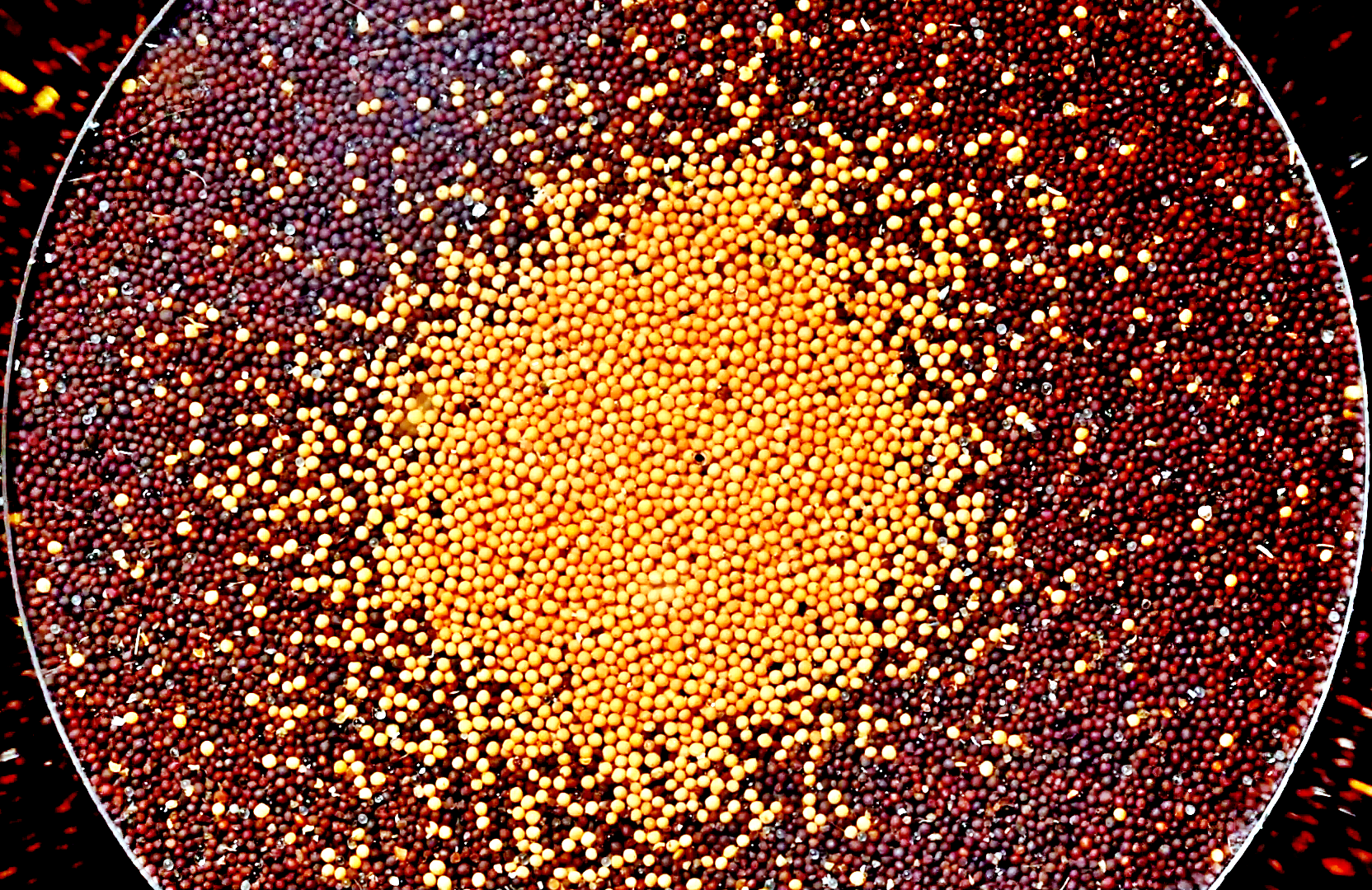}
&
\hspace{-0.17cm} \includegraphics[width=0.197\textwidth,trim={0cm 0cm 0cm 0cm},clip]{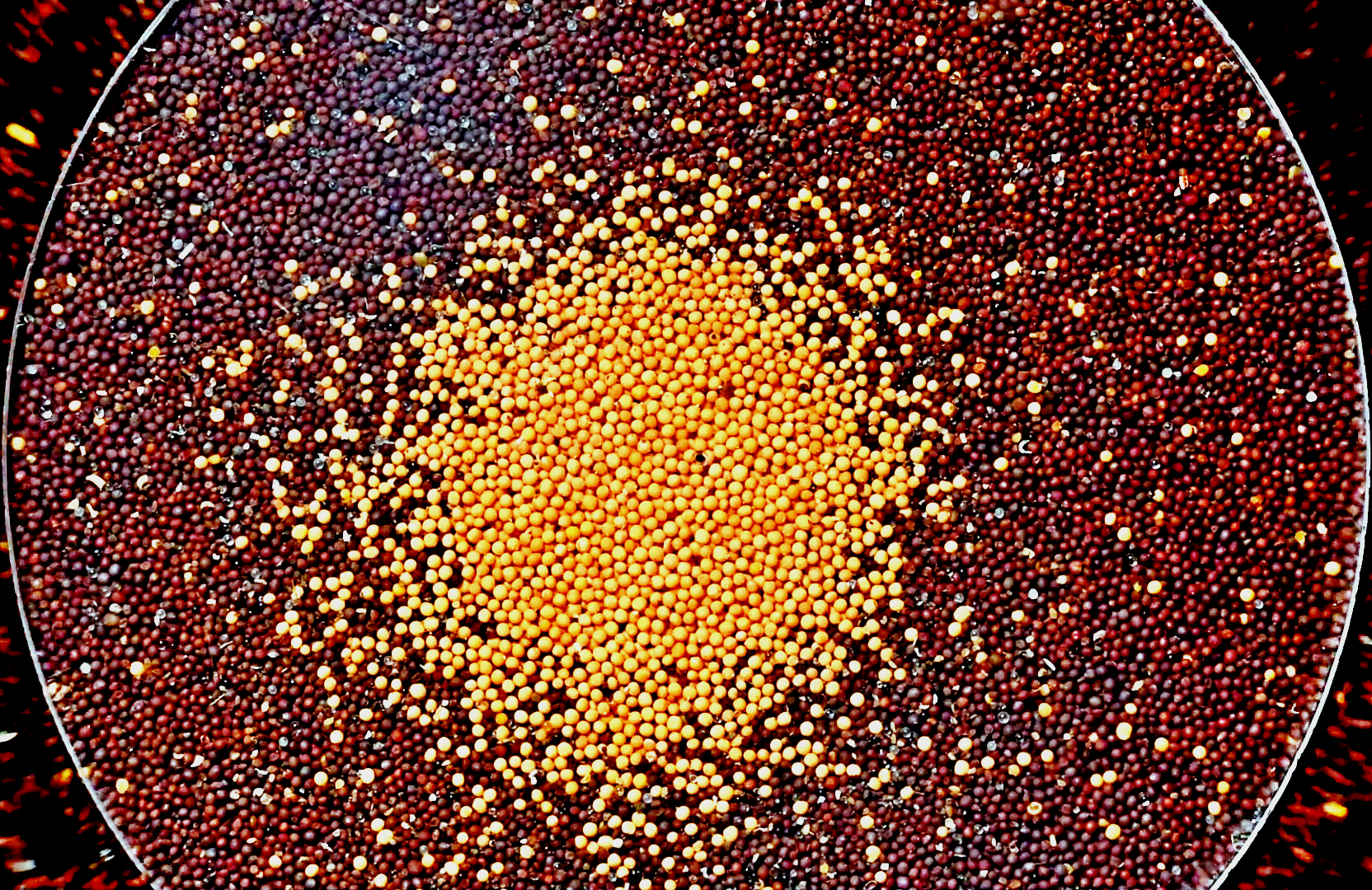}
&
\hspace{-0.17cm} \includegraphics[width=0.197\textwidth,trim={0cm 0cm 0cm 0cm},clip]{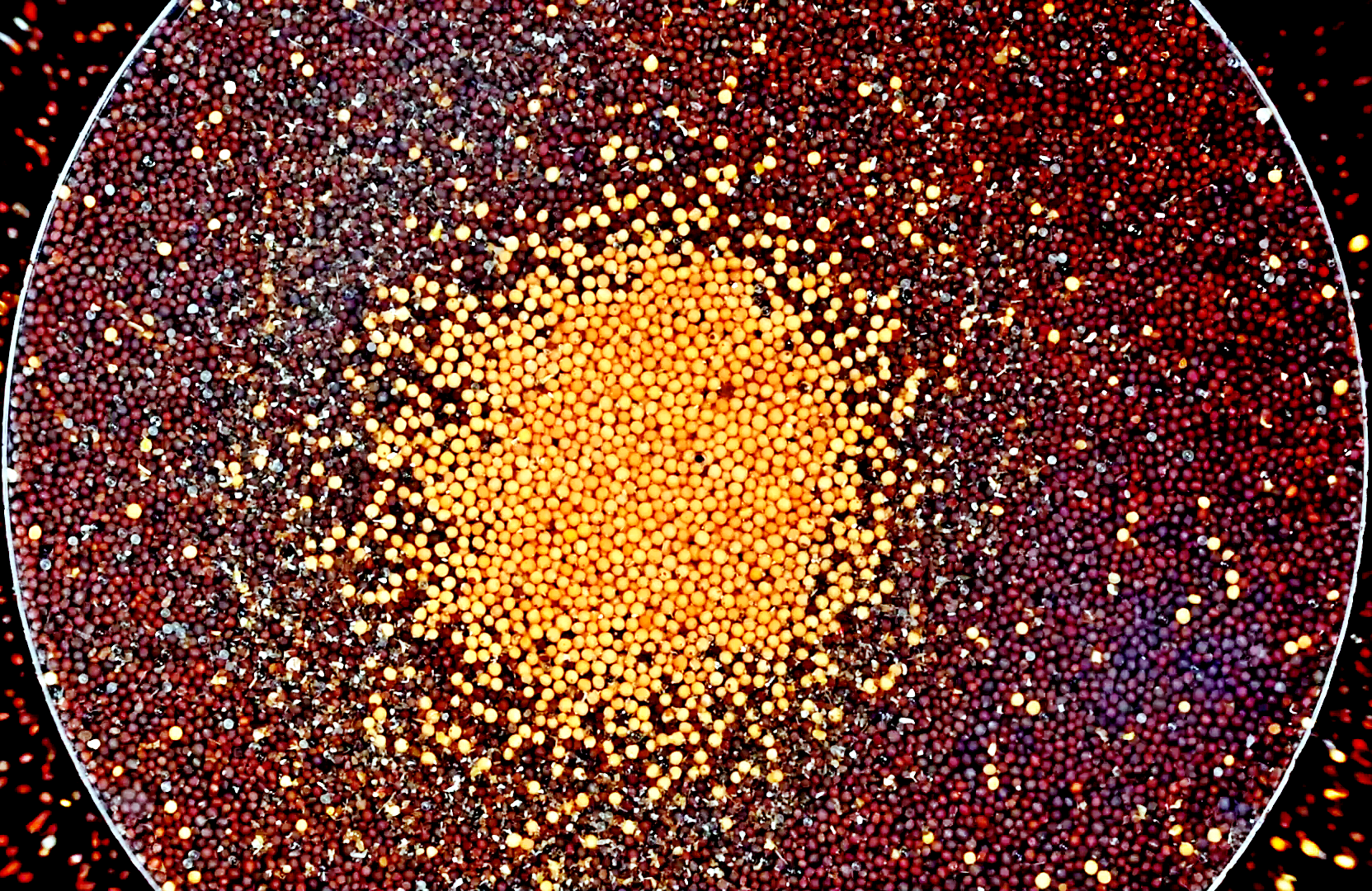}
\end{array}$
\caption{Snapshots of the transparent base holding a steady flowing grain pile. In each row, a pile is first constructed by pouring grains of one colour ($t=0$), after which grains of a contrasting colour are poured at constant rate from above.  The top row is for glass beads and the bottom row of mustard seeds.  The sequential snapshots illustrate the grains from the initial pile (black glass beads and yellow mustard seeds) progressively being displaced by the grains deposited subsequently (white glass beads and brown mustard seeds).  The mean diameter $d_p$ is $1\,$mm of the glass beads,  $1.6\,$mm of the yellow mustard seeds, and $1.35\,$mm of the brown mustard seeds.  The diameters of the base of the pile and the hopper orifice are $D = 150\,$mm and $\xi = 13\,$mm, respectively, in both the experiments.\label{fig2}}
\end{figure*}

To vividly illustrate the flow deep in the pile, we conducted a set of experiments in which a pile is first formed by grains of one colour, after which grains of a contrasting colour are dispensed from the hopper.  We used combinations of black and white glass beads, yellow and brown mustard seeds, and sand of two colours.  When viewed from the side or above, the newly deposited grains blanket the existing pile, giving the impression that they are just flowing down the surface. However, the images from the camera placed below the base plate, shown in Fig.~\ref{fig2}, show the grains in the original pile being displaced by the newly deposited grains, progressing from the periphery to the centre.  Supplementary Movies 1--3 show the radial outward motion of the particles at the base at all stages of the experiment.  Thus, the experiments provide compelling evidence of downward and radial outward motion of the particles, penetrating all the way to the base.  The rate of shrinkage of the core of the original pile decreases with time, implying that the radial velocity falls as we move inwards from the edge of the disk.  Figure~\ref{fig2} shows that over a period of 10 minutes, the black glass beads are almost entirely replaced by the white beads, but a core of the yellow mustard seeds, is still present.  The smaller radial velocity of the mustard seeds can to attributed to them being more frictional --- their angle of repose $\phi$ is $26^\circ$, compared to $22^\circ$ for glass beads.  Sand grains are even more frictional, with $\phi = 39^\circ$, and hence the radial velocity is much lower, but nevertheless the slow shrinkage of the core of the original sand pile can be discerned in Supplementary Movie 3.

\begin{figure*}[b]
\includegraphics[width=.33\textwidth]{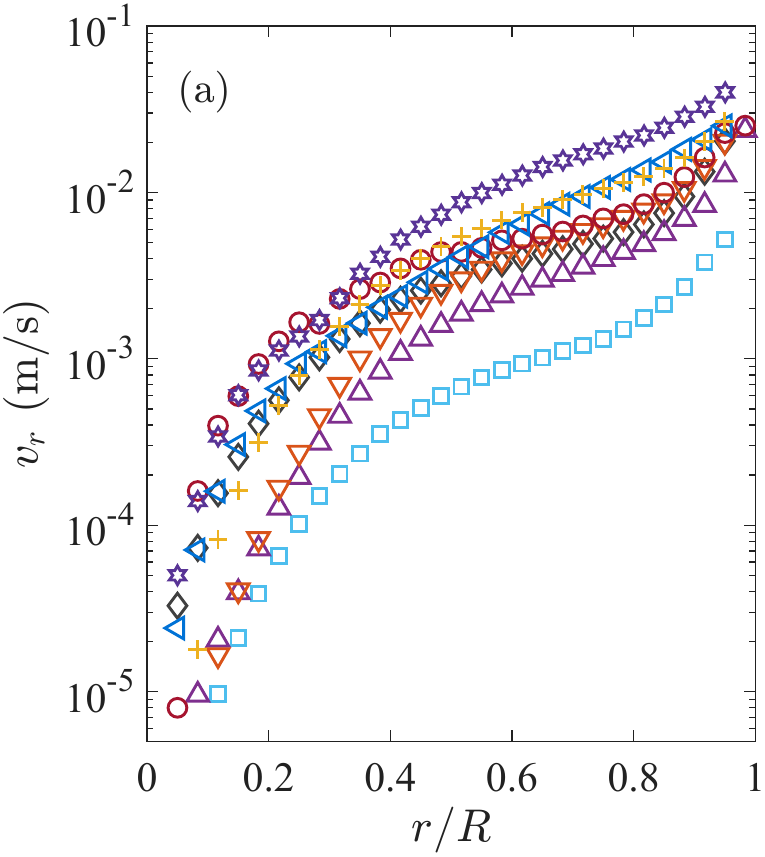}  \hspace*{5ex}  		
\includegraphics[width=.33\textwidth]{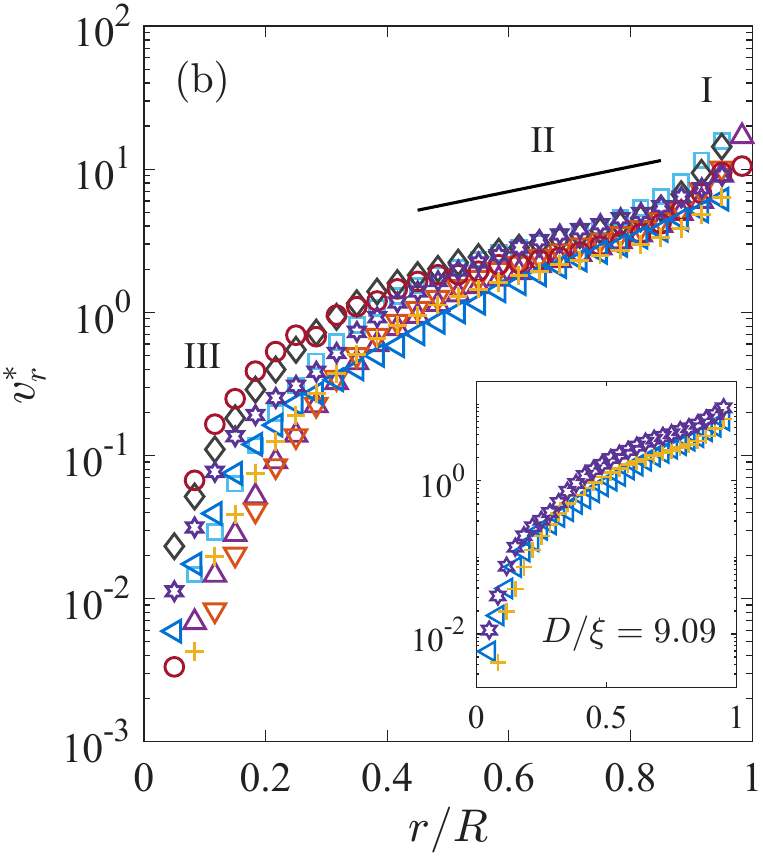}\vspace*{1em}\\
\includegraphics[width=.48\textwidth]{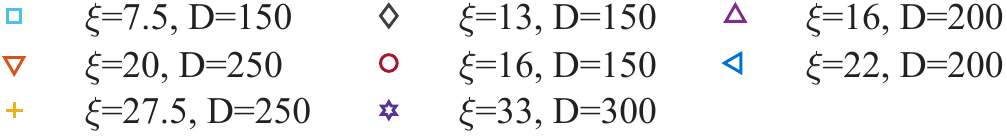}
\caption{(a) Profile of the radial velocity at the base for different values of the hopper orifice diameter $\xi$ (which sets the deposition rate $Q$) and the base diameter $D$ of the pile (see Fig.~\ref{fig1}). The radial distance $r$ is scaled by the base radius $R \equiv D/2$. (b) Profile of the normalised velocity $v_r^* \equiv \frac{v_r}{v_{\subtext{d}}} (\frac{D}{\xi})^2$, where the $v_{\subtext{d}} = Q/(\frac{1}{4}\rho \pi \xi^2)$ is the deposition velocity. The inset show very good collapse of the velocity profiles for three cases with constant $D/\xi$.  The legend gives the values of $D$ and $\xi$ in units of the particle diameter $d_p$.\label{fig3}}
\end{figure*}

The results in Fig.~\ref{fig2} and Supplementary Movies 1--3 show unambiguously the presence of flow deep in the pile, but it is also of interest to obtain quantitative measurements of the velocity field and also assess the influence of the pile dimensions, deposition rate and properties of the granular medium.  Figure~\ref{fig3}(a) shows the radial velocity profile $v_r(r)$ at the base for glass beads, obtained from the video images using particle image velocimetry (see Methods), for a wide range of $D$ and $\xi$ (which sets the deposition rate).  The velocity is measurable over several orders of magnitude before it drops to zero at the axis of symmetry.  Though the value of $v_r$ near the axis is very small, about $10^{-5}\,$mm/s, the measurements are accurate and reproducible (see Methods).  The measurement accuracy as a function of radial position is shown in Fig.~\ref{figS1}.

The scale for the velocity field is set by deposition velocity $v_{\subtext{d}} \equiv Q/(\frac{1}{4} \rho \pi \xi^2)$ from the hopper, but it can also depend on the dimensionless parameters $D/\dpart$ and $D/\xi$. When scaled by $v_{\subtext{d}}$, the velocity profiles at the base for the different pile and orifice diameters do not collapse (not shown), but we obtain a reasonable collapse of all the data when we adopt the scaling $v_r^* \equiv \frac{v_r(r)}{v_{\subtext{d}}}\, (\frac{D}{\xi})^2$ (Fig.~\ref{fig3}b). In other words, the velocity at the base varies with pile size as $D^{-2}$.  The collapse of $v_r^*$ is not entirely satisfactory --- the profiles for the different parameter sets diverge in the core of the pile --- but it is much better when the ratio $D/\xi$ is kept constant, as shown by the inset of Fig.~\ref{fig3}(b). Thus, the velocity profile depends on the fraction of the base area over which grains are deposited, which is in conformity with earlier computational evidence \cite{krishnaraj_nott2024} that the flow in a pile formed by depositing through a narrow funnel or raining uniformly are different.  We can identify three distinct regions in the scaled velocity profile in Fig.~\ref{fig3}(b): as we move radially inwards from the edge of the disk, there is first a sharp drop in $v_r$ in region I, which corresponds to the rapidly flowing surface layer impacting the base and flowing radially outward.
Further inward is the relatively wide region II in which $v_r(r)$ decreases exponentially.  Such an exponential decrease is observed frequently in shearing layers in slow (non-inertial) granular flow, and has been explained by non-local and higher gradient plasticity models \cite{mohan_etal2002, bouzid_etal2013, dsouza_nott2020}.  In region III at the core of the pile, $v_r(r)$ decays faster than exponential and reaches zero at $r=0$.

\begin{figure*}
\centering
\begin{minipage}{0.48\textwidth}
  \centering
  \includegraphics[width=\textwidth]{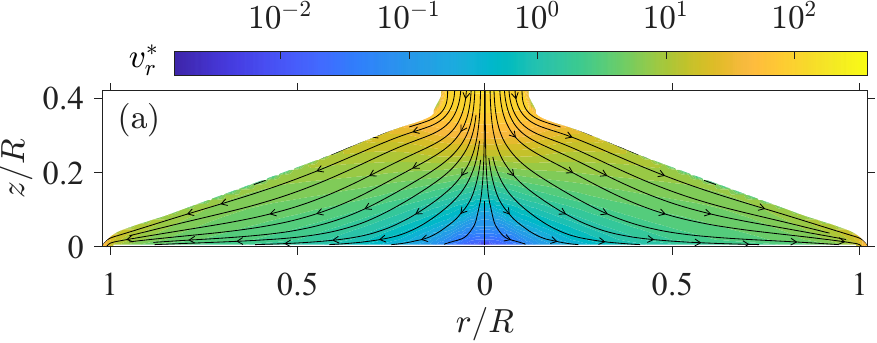}
\end{minipage}
\hspace*{1ex}
\begin{minipage}{0.48\textwidth}
  \centering
  \includegraphics[width=\textwidth]{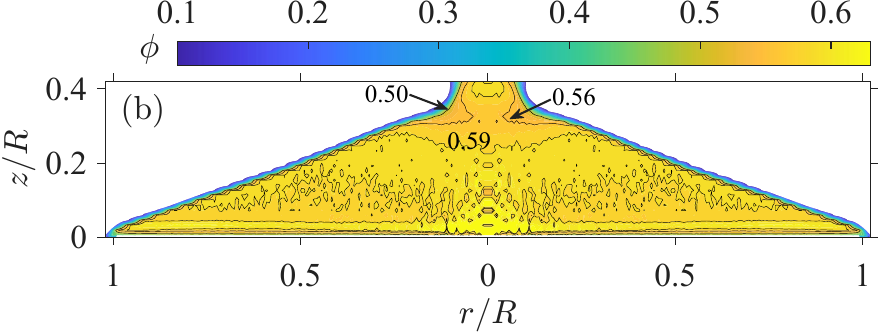}
\end{minipage}
\begin{minipage}{0.48\textwidth}
  \centering
\vspace*{1em}
  \includegraphics[width=0.64\textwidth]{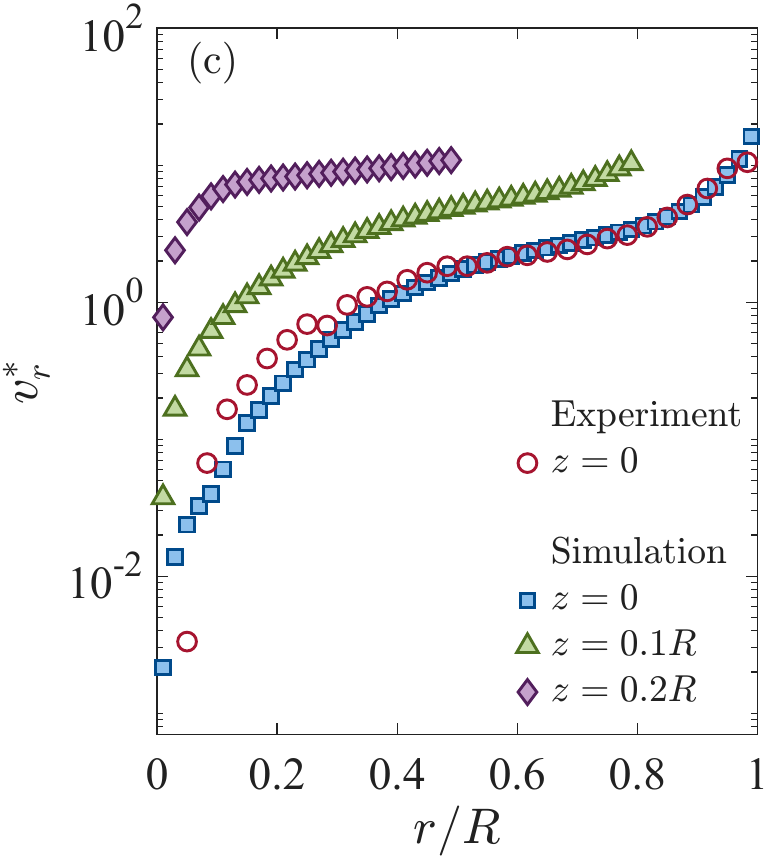}
\end{minipage}
\hspace*{1ex}
\begin{minipage}{.48\textwidth}
  \centering
  \includegraphics[width=\textwidth]{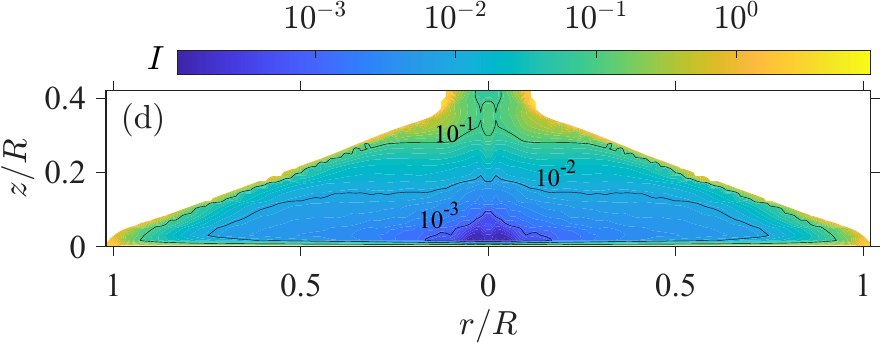}
\end{minipage}
\caption{ Results of a particle dynamics simulation for a steady-flowing pile with $D = 150\,\dpart$, $\xi = 16\, \dpart$. (a) Colour map of the scaled velocity magnitude $v^* = (v_r^2 + v_z^2)^{1/2}/v_s$ with superimposed streamlines.  (b) Colour map of the particle solids fraction $\phi$.
(c) Profiles of the radial velocity $v_r$ at the base and two horizontal planes at heights $0.1\, R$ and $0.2\, R$ above the base.  Experimental data at the base for a pile of the same dimensions are shown for comparison; we see very good agreement between the simulation and experimental data. (d) Colour map of the inertia number $I$.  We see that $I \ll 1$ except in the thin layer near the free surface and below the zone of deposition of grains at the top.\label{fig4}}
\end{figure*}

While the experimental results in Fig.~\ref{fig3} show only the flow at the base, they suggest strongly that flow should occur throughout the pile, as the velocity is expected to lowest at the base due to the retarding influence of the base plate.  To obtain an understanding of the kinematics throughout in the pile, we conducted particle dynamic simulations using the discrete element method (DEM, see Methods), a widely used computational tool that has been shown to reproduce accurately the rheology \cite{chialvo_etal2012, gautam_nott2025} and complex kinematics \cite{krishnaraj_nott2016} of granular media.  Figure~\ref{fig4}(a) displays the complete velocity within the pile; it is clear that there is no static core, and confirms the experimental observations of flow penetrating all the way till the base.  Near the base, the streamlines bear resemblance to those in a liquid jet impinging a plate -- the direction of flow varies smoothly from vertical at the symmetry axis to horizontal at the periphery.  The particle volume fraction $\phi$ is above the jamming value (loose random packing) of $\approx 0.56$ everywhere except in a thin layer at the surface (Fig.~\ref{fig4}b). The profiles of the radial velocity $v_r(r)$ at different heights from the base are shown in Fig.~\ref{fig4}(c); we see very good agreement between the simulation and experimental data at the base.  We note that the agreement is not a result of optimizing the DEM parameters, as the particle properties used in the simulations are largely taken from previous studies that simulated glass beads (see Methods).  The magnitude of the velocity increases sharply with height $z$ from the base (Fig.~\ref{fig4}a,b), and the velocity at each $z$ is maximum at the surface.

	The low volume fraction and high velocity in the surface layer is in accordance with numerous studies that have shown the surface flow to be inertial \cite{savage1979,johnson_etal1990,silbert_etal2001}.  As mentioned earlier, it is widely believed that the major part of the pile is static.  Since we observe flow throughout the pile, it is pertinent to ascertain the role of grain inertia everywhere.  A dimensionless number that characterizes the influence of inertia is the Savage number $Sa$, or equivalently the inertia number $I$,
\begin{equation}
	Sa = I^2 =  \frac{\rho_p \dot{\gamma}^2 \dpart^2}{p}
\end{equation}
where $\dot{\gamma} \equiv (2\mathbf{D}'\!\!\bm{:}\!\!\mathbf{D}')^{1/2}$ is a scalar measure of the deviatoric deformation rate tensor $\mathbf{D}'$, $p$ is the pressure, and $\rho_p$ is the intrinsic density of the particles.  The flow is in the rapid, or grain inertia, regime when $I \sim 1$, in which the stress scales as $\rho_p \dpart^2 \dot\gamma^2$; it is in the slow, or quasistatic, regime when $I \lesssim 10^{-2}$, in which the stress is independent of $\dot\gamma$.  We computed $\mathbf{D}'$ and $p$ from the continuum velocity and stress fields determined in the simulation (see Methods), and thence the spatial distribution of $I$, which is shown in Fig.~\ref{fig4}(d).  Apart from the thin layer at the surface where  $I \sim 1$ and $\phi$ is low (Fig.~\ref{fig4}b), the flow in the pile is predominantly in the quasistatic regime.

\begin{figure*}
\centering
\includegraphics[width=.33\textwidth]{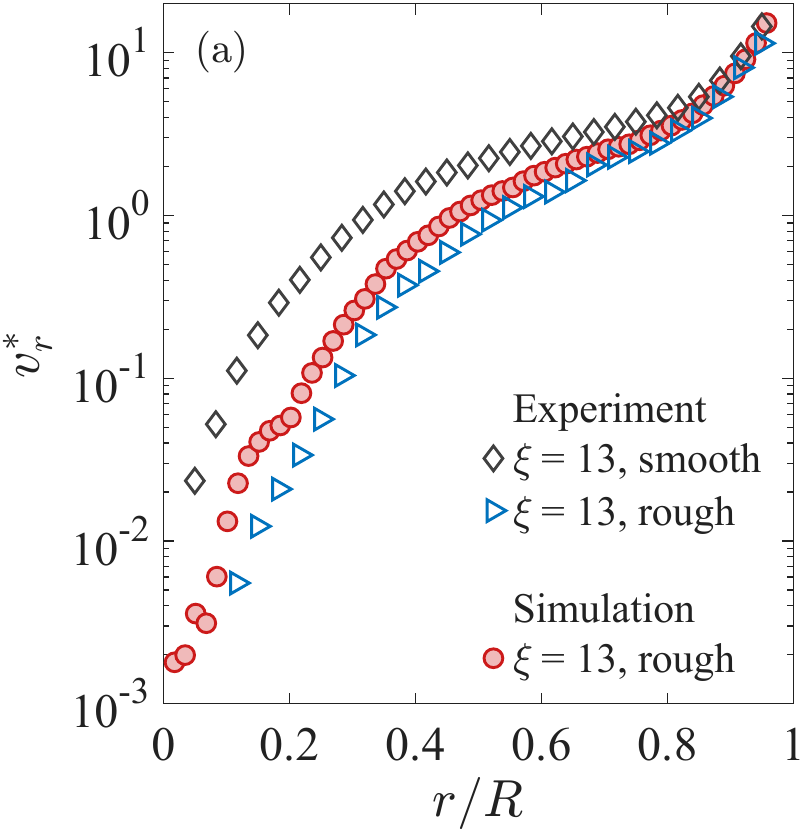} \hspace*{5ex}   \includegraphics[width=.33\textwidth]{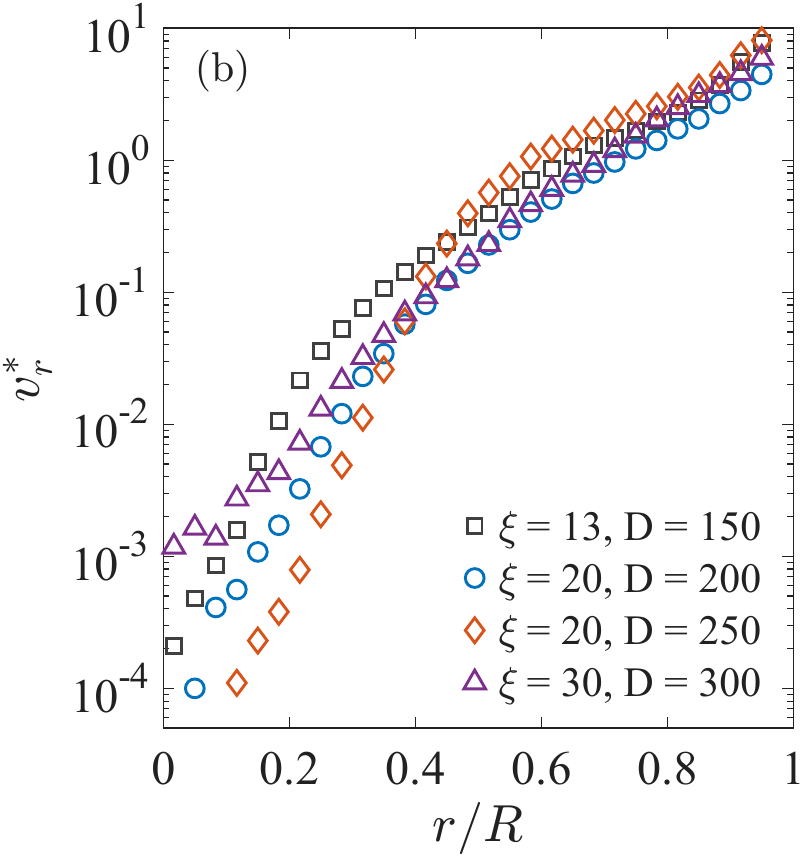} 
\caption{Effect of base friction and grain friction on the velocity profile at the base. (a) Profiles of the radial velocity at the base for piles of glass beads on smooth and rough base.  (b) Radial velocity profiles at the base in piles of mustard seeds, which are more frictional than glass beads.  The figure should be compared with Fig.~\ref{fig3}(b). \label{fig5}} \vspace{-0.4cm}
\end{figure*}

The results in Figs~\ref{fig2}-\ref{fig4} are for piles formed on a relatively smooth acrylic base.  It is well known that granular flow can be significantly affected by the roughness of boundaries. Hence, it is of interest to know how a rougher base would alter the flow in the pile.  To answer this question, we had to introduce roughness while preserving the transparency of the base.  We did so by creating spherical cap cavities (see Methods) in the acrylic disk. Particles recurrently fall into the cavities and get dislodged, having the overall effect of increasing the resistance to flow.  The roughness of the base resulted in the angle of repose of the pile of  glass beads increasing to $25^\circ$ (from $22^\circ$ for a smooth base).  Figure~\ref{fig5}(a) compares the velocity profiles for piles of glass beads formed on smooth and rough surfaces.  The velocity is substantially lower for the rough base, the reduction being more pronounced in the core of the pile, but the velocity is measurable till close to the symmetry axis.  The features of the velocity profile for the smooth and rough bases are identical, including the three regions of variation mentioned with reference to Fig.~\ref{fig3}(b).  Fig.~\ref{fig5}(a) also shows the basal velocity profile determined from a DEM simulation with a rough base (see Methods); here too we find good agreement between the experimental and simulation results (Fig.~\ref{fig5}a).  But for the lower velocity near the base, the velocity and packing fraction fields determined from the simulation with a rough base are quite similar to those in Fig.~\ref{fig4} for a smooth base.

Lastly, we show in Fig.~\ref{fig5}(b) the radial velocity profile at the base in steady flowing piles of mustard seeds for different deposition rates and base diameters.  As mentioned earlier, mustard seeds are more frictional than glass beads. The velocity profiles are qualitatively similar to those for glass beads (compare Fig.~\ref{fig3}b), apart from the decay with radial distance from the periphery being more rapid.  The radial velocity in the core of the pile is lower than that of glass beads by a factor of $\sim 10$, which manifests as a much slower shrinkage of the core of yellow mustard seeds in Fig.~\ref{fig2} and Supplementary Movie 2. Nevertheless, the data are within our measurement resolution.  For sand, whose grains are irregular in shape and have an even higher friction coefficient, the velocity in the core region is smaller than our measurement resolution.  Thus, the results in Fig.~\ref{fig5} show that increasing friction of the base or the granular medium reduces the magnitude of the velocity, but the qualitative nature of the flow field within the pile remains the same.

In conclusion, we have presented for the first time unambiguous evidence of sustained plastic flow deep inside a grain pile, refuting the widely held belief that flow is restricted to a thin layer near the free surface.  Our investigation provides new insight into the flow of granular media, and suggests that a smooth variation of velocity may also exist in the other configurations where flowing and static regions are though to coexist, such as hoppers and bunkers \cite{nedderman-book,rao_nott}.

We have investigated steady flowing piles in order to make accurate measurements of small velocities by averaging over a sufficiently long duration.  Nevertheless, we speculate that the kinematics in a pile forced by intermittent flow at the surface would be similar, because the irrelevance of inertia (Fig.~\ref{fig4}d) implies that disturbances at the surface should elicit a response everywhere in the pile if the medium is in a state of plastic yield.  Our results are therefore of relevance to geophysical phenomena and industrial applications where material is deposited intermittently on piles and heaps.  They reinforce the observations of creep deep inside quasi 2D piles \cite{deshpande_etal2021,deshpande_etal2024}, but go further by providing the full velocity field.

The presence of flow throughout the pile is consistent with recent findings that sandpiles and hill slopes are at the margin of failure \cite{roering_etal2001, ferdowsi_etal2018, deshpande_etal2021}.  Another consequence of the irrelevance of inertia in the subsurface flow is that the stress during flow must be similar to that in a static pile; Fig.~\ref{figS2} corroborates this conclusion by showing that the basal stress profile during flow is very similar to that in a static pile, with a characteristic dip under the apex \cite{jotaki_moriyama1979, smid_novosad1981}.  Consequently, our results suggest that the stress in static piles and other granular assemblies deposited under gravity is in a plastic (rather than elastic) state, possibly settling a long-standing question on the nature of stress transmission in these systems \cite{wittmer_etal1996, savage1997, savage1998}.  Our findings have substantial implications for continuum of modelling slow granular flow: the occurrence of flow throughout the pile reinforces the need for a non-local plasticity model, and the complex flow and accompanying density gradients can be used to discriminate between existing non-local models \cite{dsouza_nott2020,bouzid_etal2013,henann_kamrin2013}.  We end with the caveat is that our study was restricted to non-cohesive granular media.  We expect the flow deep inside a pile of cohesive grains, such as moist sand, to be weaker, and vanish when the cohesive force between grains is sufficiently high.
\vspace*{1.5em}

\noindent {\bf \large Materials and Methods}\vspace*{0.5em}\\

\noindent {\bf Experiments}\vspace*{0.25em}\\
Grain piles were created by depositing grains onto a transparent circular base plate of diameter $D$ from a hopper with a circular orifice of diameter $7.5 < \xi < 33\,$mm  (Fig.~\ref{fig1}a). The base plate was bolted to a rigid aluminium frame with connector bars of width $5\,$mm emerging radially from the plate.  
The circular plate with connector bars was laser-cut from a flat acrylic sheet of thickness $5\,\mathrm{mm}$. The height $h$ of the hopper orifice was 3--5$\,$mm above the virtual apex of the pile, small enough that the flow is slow (and not in gravitational free fall) but large enough that it is steady. Grains were allowed to spill over from the base (Figs~\ref{fig1}a,b) and within a short time interval, a steady flowing state is achieved where the rates of inflow from the hopper and the spill over from the base are equal.  Following an initial transient phase, the system reaches a steady flowing state in which the mass flow rate of grains supplied from the hopper balances the rate at which grains spill over from the base.  All measurements were made under this steady-state condition.

Most of the experiments were conducted with nearly spherical soda-lime glass beads of bulk density $\rho = 1417\,$kg/m$^{3}$ and a narrow size distribution centred around a mean diameter $\dpart = 1\,$mm; the static angle of repose of the pile formed by the beads was $22^\circ$. A fraction of the particles was spray painted black to conduct the displacement experiment shown in Fig.~\ref{fig2}. The painted beads exhibited a higher inter-grain friction that manifested in the angle of repose increasing to $25^\circ$.  A few experiments were performed with nearly spherical mustard seeds.  The combination of brown ($\dpart \approx 1.35\,$mm) yellow seeds ($\dpart \approx 1.6\,$mm) was used to conduct the displacement experiment shown in Fig.~\ref{fig2}. Both samples had bulk density $\rho \approx 670\,$kg/m$^{3}$ and displayed an angle of repose of $25^\circ$.  Finally, beach sand comprising grains of irregular shape was used for the displacement experiment shown in Supplementary Movie 3. An initial pile was made of yellow grains of nominal size of $\dpart \approx 1.8~\mathrm{mm}$, followed by a mixture of light brown and dark brown grains with nominal size $\dpart \approx 1.6~\mathrm{mm}$; both varieties had a density of $\rho \approx 1390\,$kg/m$^{3}$ and exhibited an angle of repose of $38^\circ$.

To study the role of base roughness on the flow,  a few experiments were conducted with an acrylic base plate that was roughened by making indentations with a hemispherical indenter. The indenter was held normal to the surface and lightly struck with a hammer to produce spherical cap cavities of diameter and depth roughly $1.5\,$mm and $0.4\,$mm, respectively.  The cavities were distributed randomly across the surface with a mean spacing of $5\,$mm. 

Particles intermittently fall into the cavities and get dislodged, having the overall effect of increasing the resistance to flow.

During flow, particles intermittently fall into the cavities and get dislodged from them, hindering the motion of other particles and increasing the resistance to flow. Importantly, the indentations leave the transparency of the base plate unaffected, thereby allowing the flow at the base to be visualized.  The increased surface friction reduced particle slip at the base, leading to an increase in the pile angle of glass beads from $21^\circ$ on a smooth surface to $25^\circ$ on the roughened surface.

The flow at the base of the pile was imaged using a video camera (Nikon D3500) positioned below the transparent base (Fig.~\ref{fig1}a). The images were recorded at 60 frames per second at a resolution of 1920$\times$1080 pixels.  The instantaneous velocity field was obtained from two adjacent frames using the particle image velocimetry toolbox PIVlab~\cite{thielicke_sonntag2021}.  A smoothly varying velocity field was then obtained by averaging over time and, with the assumption of axisymmetric flow, spatially in annular bins of width $\Delta r = R/30$.  This process yields the radial velocity profile $v_r(r)$ were used to characterize the steady radial outflow.

PIV of dense granular flows is known to yield sub-pixel accuracy in displacement when averaged over a sufficiently large number of image pairs
thereby yielding accurate velocity measurements. We obtained an overall estimate of measurement accuracy from multiple measurements of the velocity profile, each averaged in the annular bins corresponding to every radial location and over a time interval of $90\,$s.  The standard deviation from these measurements is indicated by the error bars in Fig.~\ref{figS1} -- we see that the error is comparable with the mean velocity only close to the symmetry axis $r=0$, where $v_r$ vanishes, and the errors are much smaller than the mean away from  the axis.

\vspace{1em}

\noindent {\bf Simulations}\vspace*{0.25em}\\
The particle dynamics simulations were conducted using the discrete element method (DEM), in which particles are treated as deformable, and their interactions are dissipative and frictional.  We used the open source DEM package LIGGGHTS~\cite{kloss_etal2012} to  carry out the simulations.  DEM has been elaborated in many previous studies \cite{cundall_strack1979,silbert_etal2001,santos_etal2020}, hence we give only a brief description here. Consider a  particle assembly where particle $i$ has diameter $d_i$, is centred at position vectors $\mathbf{x}_i$ and has translational and rotational velocities ($\mathbf{v}_i$, $\boldsymbol{\omega}_i $).  For particles $i$ and $j$ in contact, rather than compute their deformations in detail, the interaction force and torque are determined from their `overlap' $\delta \equiv \frac{1}{2} (d_i + d_j) - |\mathbf{x}_{ij}|$ and the relative velocity at the point of contact.  We used the Hertz-Mindlin contact model \cite{johnsonkl} with viscous damping and rolling friction (model `hertz' with rolling friction `epsd2' in LIGGGHTS), in which the normal and tangential components of the force on particle $i$ are 

\begin{eqnarray*}
	\mathbf{F}_{\mathrm n} & = & \frac{\sqrt{2} E}{3(1-\nu^2)} \, \overline{d}^\frac{1}{2} \delta^\frac{3}{2} \mathbf{n} - \overline{m} \, \gamma_{\mathrm n} \mathbf{v}_{\mathrm n},\\
	\mathbf{F}_{\mathrm t} & = & \left\{ \begin{array}{ll}
					-\Frac{\sqrt{2} E}{(2 - \nu)(1 + \nu)} \, \overline{d}^\frac{1}{2} \delta^\frac{1}{2} \mathbf{s} \; - \overline{m}\, \gamma_{\mathrm t} \, \mathbf{v}_{\mathrm t}, \;\; & \text{if}\; |\mathbf{F}_{\mathrm t}/\mathbf{ F}_{\mathrm n}| < \mu\\[8pt]
					- \mu\, |\mathbf{F}_{n}| \, \mathbf{ v}_{\mathrm t}/|\mathbf{ v}_{\mathrm t}|, & \text{otherwise}.
\end{array} \right.
\end{eqnarray*}
and the torque due to rolling friction is \cite{ai_etal2011} 
\begin{eqnarray*}
\mathbf{T} & = &
\begin{cases}
\dfrac{E \, \overline{d}^\frac{5}{2} \delta^\frac{1}{2}}{2^\frac{3}{2} (2 - \nu)(1 + \nu)} \boldsymbol{\theta},
& \text{if } |\mathbf{T}_r/(F_n \overline{d})|
< \frac{1}{2} \mu_r, \\[8pt]
\frac{1}{2} \mu_r F_n \overline{d} \dfrac{\boldsymbol{\theta}}{\left\| \boldsymbol{\theta} \right\|},
& \text{otherwise}.
\end{cases}
\end{eqnarray*}
Here, $\mathbf{v}_n$ and $\mathbf{v}_t$ are the normal and tangential components, respectively, of the relative velocity of particle $i$ with respect to $j$ at the point of contact, $\mathbf{s} \equiv \int_0^\tau \mathbf{v}_{\mathrm t}\, \dup t'$ and $\boldsymbol{\theta} \equiv \int_0^\tau (\boldsymbol{\omega}_i - \boldsymbol{\omega}_j)\, \dup t'$ are the tangential and angular displacements during contact, and $\overline{d}$, $\overline{m}$ are half the harmonic mean of the diameter and mass,  i.e.\ $\overline{d} \equiv (1/d_i + 1/d_j)^{-1}$. The parameters characterizing the interaction are the Young's modulus $E$, Poisson ratio $\nu$, coefficient of restitution $e$ (which determines the damping factors $\gamma_n$, $\gamma_t$~\cite{tsuji_etal1992}), Coulomb friction coefficient $\mu$, and rolling friction coefficient $\mu_r$ which are assumed to be the same for all the particles.  The motion of each particle is determined by integrating Newton's second law, assuming pairwise additivity of the interaction forces.

	To prevent crystalline order, an equal mixture of grains of diameters $0.94\, \dpart$ and $1.06\, \dpart$ was used, with mean $\dpart = 1\,$mm.  As the simulations were to be compared with the experiments on glass beads, the physical and interaction properties of the particles were largely taken from previous studies~\cite{silbert_etal2001,krishnaraj_nott2016,zhou_etal2002} that simulated glass beads:  $\rho_p = 2500\,$kg/m$^3$, $\nu = 0.3$, $e = 0.71$, $\mu = 0.55$, $\mu_r = 0.04$.  The only major deviation was that the Young's modulus used, $E = 10^7\,$Pa, is smaller by a factor of $10^3$ from that of glass; this is commonly done in DEM simulations to reduce the time-step of integration and thereby computational time.  Nevertheless, the chosen value of $E$ is sufficiently large that the overlap is very small ($\delta < 10^{-3} \dpart$) and the macroscopic behaviour mimics that of hard glass beads.  Additionally, the friction coefficients $\mu$ and $\mu_r$ were chosen after tuning slightly to match the static angle of repose with the experimental measurement.  To mimic the smooth acrylic base plate in the experiment, the friction coefficients for grain-wall interactions were lower at $\mu = 0.35$ and $\mu_r = 0.035$.  To simulate the experiment with a rough base (Fig.~\ref{fig5}a), $\mu_w$ was set to 0.55, but all other parameters were kept the same.	

	Simulation of a pile of base diameter $D = 150\, \dpart$ required $\approx 3 \times 10^5$ particles.  Particles exiting the periphery of the pile at the base were removed from the simulation, and new particles were added at the top to keep the number roughly constant.  The velocity, packing fraction and stress fields were determined by averaging over time and in annular bins of width $\Delta r = 1\,\dpart$, as in the experiments.  To determine the variation of the inertia number $I \equiv  \rho_p^\frac{1}{2} \dot{\gamma} \dpart/p^\frac{1}{2}$ in Fig.~\ref{fig4}(d), the $\dot{\gamma}$ and pressure fields were determined from the velocity and stress fields using their definitions $\dot{\gamma} = \left(2\left[(\frac{\partial v_r}{\partial r})^2 +  (\frac{v_r}{r})^2 + (\frac{\partial v_z}{\partial z})^2\right] + \left(\frac{\partial v_r}{\partial z} + \frac{\partial v_z}{\partial r} \right)^2 \right)^\frac{1}{2}$, $p = -(\sigma_{xx} + \sigma_{yy} + \sigma_{zz})/3$.

\vspace*{1em}

\bibliographystyle{unsrt}
\bibliography{granflow_refs}

\section{Extended data}

\noindent {\bf Measurement accuracy of the velocity}

\begin{figure}[h]
\includegraphics[width=0.33\textwidth]{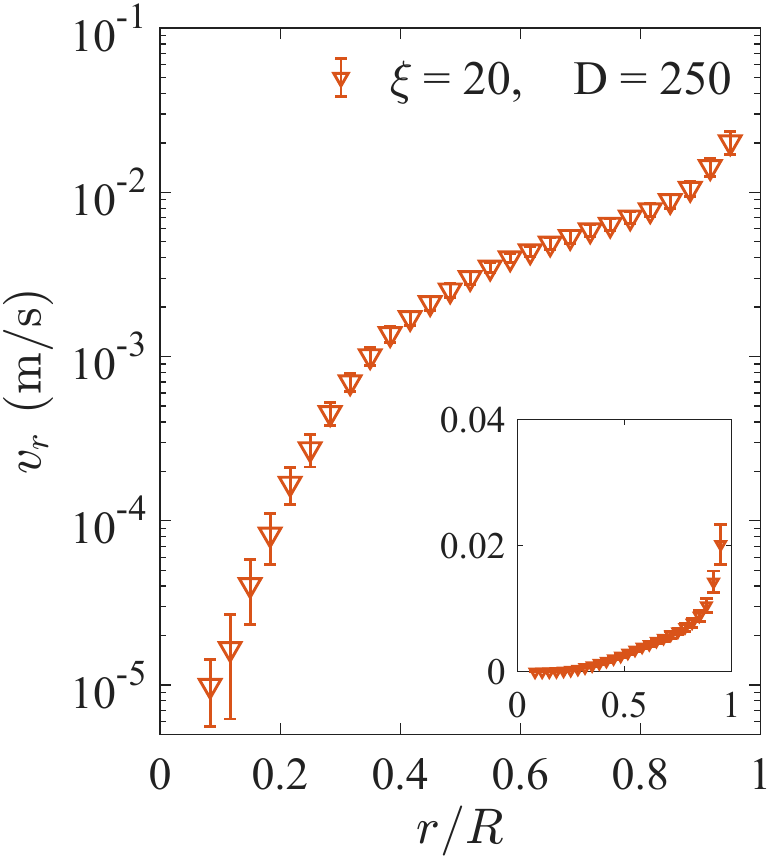}
\caption{Estimation of the accuracy of the measurements of the radial velocity $u_r$ at the base from multiple measurements.  The legend gives the diameters of the pile and hopper orifice in units of $d_p$.  The length of the error bar for each measurement is twice the standard deviation from the multiple measurements.\label{figS1}}
\end{figure}

	Determination of velocity by particle image velocimetry suffer from several sources of error, some inherent in the measurement technique, including pixel and colour (or grayscale) resolution of the video camera, the speed of capture, and uniformity of lighting.  In addition, the physical system under study itself has inherent fluctuations.  We obtained an overall estimate of measurement accuracy, which consolidates all the above-mentioned potential sources of error, from multiple measurements of the velocity profile, each averaged in the annular bins corresponding to every radial location and over a time interval of $90\,$s.  Figure~\ref{figS1} shows the result of such an exercise: the symbols represent the mean from all the measurements and the length of each error bar equals twice the standard deviation.  It is clear that the errors are significant compared to the mean close to the axis of symmetry ($r = 0$), where the mean velocity is zero, but are much smaller than the mean away from the axis. 

\noindent {\bf Basal stress during flow}

\begin{figure}
 \includegraphics[width=0.33\textwidth]{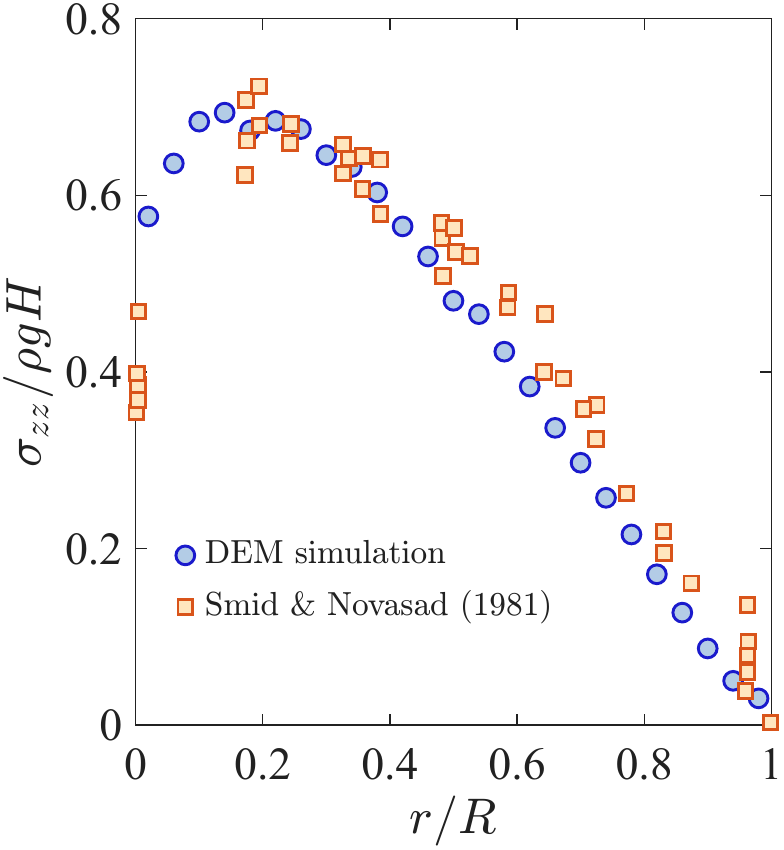} \vspace*{1em}
\caption{Profile of the vertical normal stress at the base of a steady flowing pile, obtained from the DEM simulation that yielded the velocity field in Fig.~\ref{fig4}, compared with data of Ref.~\cite{smid_novosad1981} for static piles of sand.  The stress is scaled by the hydrostatic head $\rho g H$ at the symmetry axis, where $H$ is the height of the pile's virtual apex.\label{figS2}}
\end{figure}

	In addition to particle velocities, the DEM simulations provide the contact forces on every grain, from which the stress tensor may be computed \cite{silbert_etal2001,krishnaraj_nott2024}.  Figure~\ref{figS2} shows the variation of the vertical normal stress $\sigma_{zz}$ at the base of the pile gleaned the simulation for which the velocity field is displayed in Fig.~\ref{fig4}.  The data of Ref.~\cite{smid_novosad1981} for  static piles of sand is shown for comparison.  The well-known dip in $\sigma_{zz}$ at the axis of symmetry is evident.

\newpage

\end{document}